\documentclass[12pt]{article}
\usepackage{amsmath}
\usepackage{graphicx,psfrag,epsf}
\usepackage{enumerate}
\usepackage{amsfonts}
\usepackage{amssymb}
\usepackage{psfrag}
\usepackage{enumerate}
\usepackage{xcolor}
\usepackage{soul}
\usepackage{natbib}
\usepackage{url} 

\newcommand{\blind}{1}

\begin{document}

\def\spacingset#1{\renewcommand{\baselinestretch}%
{#1}\small\normalsize} \spacingset{1}

\def\trans{^{\top}}
\def\cov{\hbox{cov}}
\def\var{\hbox{var}}
\def\log{\hbox{log}}
\def\trace{\hbox{trace}}
\def\bs{\boldsymbol}
\def\mbf{\mathbf}

\def\balpha{\bs{\alpha}}
\def\bbeta{\bs{\beta}}
\def\btheta{\bs{\theta}}
\def\bmu{\bs{\mu}}
\def\bnu{\bs{\nu}}
\def\bOmega{\mbf{\Omega}}

\def\bDa{\mbf{D}_{\balpha}}
\def\bDb{\mbf{D}_{\bbeta}}
\def\bC{\mbf{C}}
\def\bY{\mbf{Y}}
\def\bZ{\mbf{Z}}
\def\bB{\mbf{B}}
\def\bBy{{\mbf{B}_{y,}}}
\def\bBz{{\mbf{B}_{z,}}}
\def\bBy_i{{\mbf{B}_{y,i}}}
\def\bBz_i{{\mbf{B}_{z,i}}}
\def\bV{\mbf{V}}
\def\bThmu{\mbf{\Theta}_{\bmu}}
\def\bThnu{\mbf{\Theta}_{\bnu}}
\def\bthmu{\bs{\theta}_{\bmu}}
\def\bthnu{\bs{\theta}_{\bnu}}
\def\bThf{\mbf{\Theta}_\mbf{f}}
\def\bThfj{\mbf{\Theta}_{\mbf{f} j}}
\def\bThg{\mbf{\Theta}_\mbf{g}}
\def\bThgj{\mbf{\Theta}_{\mbf{g} j}}
\def\bSigab{\mbf{\Sigma}_{\balpha\bbeta}}
\def\bSigma{\mbf{\Sigma}}
\def\bSigabi{\mbf{\Sigma}_{\balpha\bbeta,i}}

\if1\blind
{
  \title{\bf Robust Joint Modelling of Sparsely Observed Paired Functional Data}
  \author{Huiya Zhou, Xiaomeng Yan, and Lan Zhou}
  \maketitle
} \fi

\if0\blind
{
  \bigskip
  \bigskip
  \bigskip
  \begin{center}
    {\LARGE\bf Robust Joint Modelling of Sparsely Observed Paired Functional Data}
\end{center}
  \medskip
} \fi

\bigskip
\begin{abstract}
A reduced-rank mixed effects model is developed for robust modelling of sparsely observed paired functional data. In this model, the curves for each functional variable are summarized using a few functional principal components, and the association of the two functional variables is modelled through the association of the principal component scores. 
A multivariate scale mixture of normal distributions is used to model the principal component scores and the measurement errors in order to handle outlying observations and achieve robust inference. The mean functions and principal component functions are modelled using splines, and roughness penalties are applied to avoid overfitting. An EM algorithm is developed for computation of model fitting and prediction. 
A simulation study shows that the proposed method outperforms an existing method, which is not designed for robust estimation. The effectiveness of the proposed method is illustrated through an application of fitting multi-band light curves of Type Ia supernovae.
\end{abstract}

\noindent%
\vfill

\newpage
\spacingset{1.45} 

\section{Introduction}\label{sec1}
This paper is motivated by a need in astrostatistics to study the evolution of Type Ia supernovae (SNeIa) through empirically quantifying the shapes of Type Ia supernova light curves and quantitatively relating the shape parameters with the intrinsic properties of SNeIa.
SNeIa are ``standardizable" candles in cosmology, widely used to measure the expansion rate of the Universe. Light curves are measurements of brightness as a function of time. By measuring the photon flux through different astronomical filters, light curves of different energy bands are measured, including the U (ultraviolet) band, R (red) band, I (infrared) band, etc. 
One of the remarkable features of SNeIa is their homogeneous nature, and as a result, the luminosity evolution dispersion is low. 
However, SNeIa consist of a considerable amount of peculiarities with differing light curve shapes, adding more heterogeneity and complexity to the light curve modelling. Peculiar SNeIa include SN1991bg-like ones which are known to have faster decline rate after the maximum brightness, to show less or no sign of right shoulders in the R band light curves, and to lack the secondary peak in the I band light curves compared with normal SNeIa \citep{filippenko1992subluminous}. 
The SN1991bg-like SNeIa are manually removed from the training sample in some existing light curve fitting models, e.g. \cite{guy2007salt2}, to mitigate the disturbing effect.

\begin{figure}[hbt]
\centering
\includegraphics[width=0.9\textwidth]{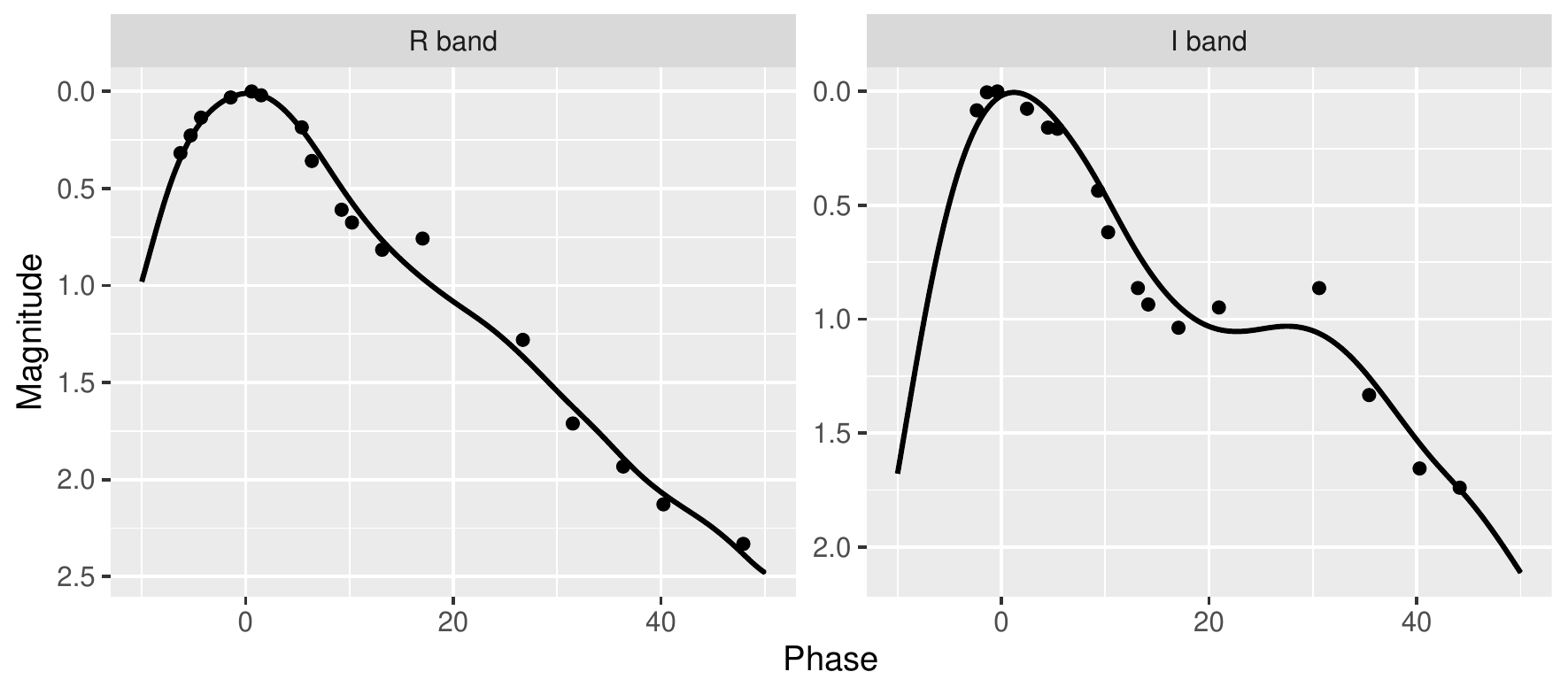}
\caption{Observed and fitted R band and I band light curves of SN2005iq. The points are light curve observations. The lines are the fitted curves using the method of \cite{zhou2008joint}. }\label{fig:motivation}
\end{figure}

Since observed SNeIa light curves can be viewed as sparsely observed functional data
\citep{he2018characterization}, one can apply the method of functional principal component analysis for paired functional data developed by 
\cite{zhou2008joint} to jointly 
model data of two energy bands simultaneously. This method is based on a reduced-rank mixed effects model for paired functional data, where splines are used to model the principal component (PC) functions and PC scores are
modelled as normally distributed random effects. 
Figure~\ref{fig:motivation} shows the fitted light curves by applying this method on the observed R band and I band data for the Type Ia supernova SN2005iq. 
While the fitting for the R band observations is reasonably good, the fitted light curve in I band tends to be flatter around the second peak. The second peak has a physical meaning for this type of supernova. 
The reason that the method missed the second peak in the I band is that it is not resistant to outlying observations, i.e., nonignorable peculiarities in SNeIa light curve observations.
This motivates us to develop a robust functional PCA method that can jointly model observations from two energy bands and is resistant to outlying observations. 

\cite{lange1989robust} suggested that the Student $t$ distribution provides a powerful tool
for handling outliers in a wide range of settings. Inspired by this work, we propose to extend the mixed-effects FPCA model of 
\cite{zhou2008joint}
by using the Student $t$ distribution to
model the PC scores and the measurement errors to obtain robustness of the
methodology. More generally, our framework incorporates a broader class of
distributions for modelling the PC scores, i.e., the scale mixture of normal (SMN) distributions
\citep{andrews1974},
for which the Student $t$ distribution is a special case.
By treating the scale parameter in the hierarchical representation of 
SMN as a latent variable, we develop an EM algorithm to estimate parameters in our robust reduced-rank mixed
effects model for paired functional data.

Several methods have been developed for robust modelling of single
curve functional data. When the entire function is observed, \cite{locantore1999robust} proposed an approach
that projects the data onto a sphere or an ellipse around a robust estimate
of the center of the data and then performs the usual PCA on the 
projected data.
\cite{gervini2008robust} extended the work of \cite{locantore1999robust},
introduced the concepts of functional median and functional spherical PCs, and studied the robustness 
properties of the approach. 
\cite{hyndman2007robust}, \cite{hyndman2009forecasting},
and \cite{bali2011robust} proposed a projection 
pursuit (PP) approach for robust functional PCA.
\cite{gervini2009detecting} used splines to model the PC
weight functions, and modelled the functional PC scores 
using fat-tailed distributions to achieve robustness. 
\cite{sawant2012functional}  used robust functional PCA for functional 
outlier detection. 
\cite{boente2015s} 
proposed a new class of
estimators for PCs based on robust scale estimators.

With the exception of \cite{gervini2009detecting}, all robust methods
mentioned above for single curve functional data require that entire
functions are observed, therefore they are not directly applicable to
the sparsely observed functional data that we encountered in our application.
While these works have focused on functional data of single curves, this paper develops 
a method for robust modelling of paired curves, which, to the best of our knowledge, is new in
functional data analysis. Our joint modelling
approach allows strength borrowing through modelling the correlation of the
paired curves and thereby improves statistical efficiency. Treating
the scale parameter of SMN as a latent variable in an EM algorithm for robust
estimation is also a novel approach in the functional data literature. 

There is a related literature of functional canonical correlation analysis that can be used to study the correlation of paired functional data. In particular, by extending \cite{leurgans1993canonical}, 
\cite{boente2022robust} 
developed a robust smoothed canonical correlation analysis method for functional data. But that method assumes that the whole functions are observed and does not address the issue of only having sparse observations of the functions. In contrast, our method has the advantage of being able to estimate individual functions or curves with a small number of observations by borrowing strength across curves. 

The rest of the paper is organized as follows. Section \ref{sec:model} gives specifics of the proposed model. Section \ref{sec:estimation} develops the estimation method and
computation algorithm, and addresses model selection issues.
Section \ref{sec:simu} presents results from a simulation study. Section \ref{sec:AIDS} applies the proposed method to
a real Type Ia supernova dataset
and compares it with the method of \cite{zhou2008joint}. 
R code implementing the proposed method can be found on Github (https://github.com/freedom00y/pairedfda\_code).

\section{Reduced-rank mixed effects models with scale mixture of normals}\label{sec:model}

Let $Y(t)$ and $Z(t)$ denote the measurements of two separate
functional variables at time $t$, where $t$ is in a finite interval $I \subset \mathbb{R}$. 
\cite{zhou2008joint} proposed the following reduced-rank mixed-effects (RRME) model 
\begin{eqnarray}\label{eq:mixed-effects}
\begin{aligned}
Y(t) & = \mu(t) + \sum_{j=1}^{k_\alpha} f_j(t) \alpha_{j} + \epsilon(t)
= \mu(t) + \mathbf{f}(t)\trans \balpha  + \epsilon(t), \\
Z(t) & = \nu(t) + \sum_{j=1}^{k_\beta} g_j(t) \beta_{j} + \xi(t)
= \nu(t) + \mathbf{g}(t)\trans \bbeta + \xi(t),
\end{aligned}
\end{eqnarray}
where $\mu(t)$ and $\nu(t)$ are the mean functions for the two functional
variables, $\mbf{f}(t)=(f_1(t), f_2(t), \dots, f_{k_\alpha}(t))\trans $ and
$\mbf{g} = (g_1(t), g_2(t), \dots, g_{k_\beta}(t))\trans $ are vectors of
PC functions, $\balpha=(\alpha_1,\ldots,\alpha_{k_\alpha})$ and $\bbeta=(\beta_1,\ldots,\beta_{k_\beta})$ are subject
specific vectors of PC scores and are the random effects, $\epsilon(t)$ and
$\xi(t)$ are subject specific measurement errors.  In
\cite{zhou2008joint}, the PC scores $\balpha$, $\bbeta$ and the error
terms $\epsilon(t)$, $\xi(t)$, are all assumed to follow normal distributions. 
The normality assumption makes the model fitting of \eqref{eq:mixed-effects} not
resistant to the influence of outlying data points.

To introduce robustness in the RRME model, we propose to replace the normal
distribution by a distribution with fat tails. The scale mixture of
normal (SMN) distributions introduced by 
Andrews \& Mallows (1974)
will serve our purpose. We say that $X \sim {\rm SMN} (\mu,\phi;\mathcal{H})$
with position parameter $\mu \in \mathbf{R}$, scale parameter $\phi >
0$ and mixing distribution $\mathcal{H}$, if it can be written as $X
\overset{d}{=} \mu + U^{-1/2}X_0$, where $X_0 \sim
\mathcal{N}(0,\phi)$, and $U$ is a positive random variable with
distribution $\mathcal{H}$. Alternatively, $X$ has the following hierarchical representation:
\begin{eqnarray}\label{eq:smnhier}
X|U=u\sim \mathcal{N}(\mu,\phi/u),\quad U\sim \mathcal{H}.
\end{eqnarray}
When $\mathcal{H}$ is a point mass distribution concentrated at point $1$, ${\rm SMN}(\mu,\phi;\mathcal{H})$
reduces to  $\mathcal{N}(\mu,\phi)$. When
$\mathcal{H}=\mathcal{H}_\gamma= \text{Gamma}(\gamma/2,\gamma/2)$, $X$ follows a generalized Student
$t$ distribution with degrees of freedom $\gamma$, mean $\mu$ and scale parameter
$\phi$~\citep{theodossiou1998financial}. When
$\mathcal{H}=\mathcal{H}_\gamma=  \text{Beta}(\gamma,1)$, $X$ follows a slash
distribution with degrees of freedom $\gamma$
\citep{osorio2016influence}. Generally speaking, both generalized Student $t$ distributions and slash
distributions have fatter tails than normal distributions, making
them suitable to model outlying observations.
When $\gamma\rightarrow \infty$, both
Gamma$(\gamma/2,\gamma/2)$  and Beta$(\gamma,1)$ will converge to
the point mass at 1, and it follows that both the generalized Student $t$ and slash
distributions converge to a normal distribution.

Similar to \eqref{eq:smnhier}, let $U \sim \mathcal{H}$ be a
positive valued latent variable, where $\mathcal{H}$ is a
probability distribution. We assume that, given $U=u$,
the PC scores
$(\balpha\trans, \bbeta\trans)\trans$ and the measurement errors
$\epsilon(t), \xi(t)$ are independent for all $t \in I$ and have conditional
distributions
\begin{eqnarray}\label{eq:slash-model}
\begin{pmatrix}\balpha\\ \bbeta\end{pmatrix}\bigg  \vert U = u 
\sim \mathcal{N} \biggl(0, \frac{1}{u}\bSigab\biggr),\quad \bSigab = 
\begin{pmatrix}
\bDa & \bC\\
\bC\trans & \bDb\\
\end{pmatrix};
\end{eqnarray}
\begin{eqnarray}\label{eq:error}
  \epsilon(t) \vert U=u \sim N\biggl(0, \frac{\sigma^2_\epsilon}{u}\biggr),
  \quad \xi(t) \vert U=u \sim  N\biggl(0, \frac{\sigma^2_\xi}{u}\biggr),\quad t \in I.
\end{eqnarray}
The submatrices of $\bSigab$ can be rephrased as conditional
covariance matrices, i.e.,
$\var(\balpha_i|U=u) = \bDa/u $, $\var(\bbeta_i|U=u) = \bDb/u$, and $
\cov(\balpha_i,\bbeta_i|U=u) = \bC/u$. 
Equations~\eqref{eq:mixed-effects}, \eqref{eq:slash-model} and
\eqref{eq:error} together specify our robust reduced-rank mixed effects model for
paired functional data.

If $U\sim \text{Gamma}(\gamma/2,\gamma/2)$, then $(\balpha\trans,
\bbeta\trans)\trans, \epsilon(t), \xi(t), \bY(t), \bZ(t)$, for all $t \in I$,
follow generalized Student $t$ distributions with degrees of freedom
$\gamma$, and we call the reduced-rank mixed-effects model RRME-t
model. If $U\sim \text{Beta}(\gamma,1)$, then $(\balpha\trans, \bbeta\trans)\trans, \epsilon(t), \xi(t), \bY(t), \bZ(t)$, for all $t \in I$, follow slash distributions with degrees of freedom $\gamma$, and we call the reduced-rank mixed-effects model RRME-slash model.
When $\mathcal{H}$ is a point mass distribution focusing on the value
$1$, we return to the model in \cite{zhou2008joint}, which we call RRME-normal model.

For identifiability, the PC functions are subject to the orthogonality constraints $\int f_kf_l=\delta_{kl}$
and $\int g_k g_l=\delta_{kl}$, with $\delta_{kl}$ being the Kronecker delta, i.e.,
$\delta_{kl} = 1$ if $k=l$, and $0$ otherwise.
Moreover, given $U = u$, the PC scores
$\alpha_{j},j\in\{1,\dots, k_\alpha\}$, are conditionally independent with
decreasing variances and the PC scores
$\beta_{j}, j\in\{1,\cdots,k_\beta\}$, are also conditionally independent with
decreasing variances. 

To estimate unknown functions in the robust reduced-rank mixed effects
model, we follow the same idea as that in \cite{zhou2008joint} and represent
them as spline functions and reduce the problem to estimation of
spline coefficients. 
Specifically, we represent the mean functions $\mu, \nu$ and elements
of $\mbf{f}$ and $\mbf{g}$
 as members of the same space of spline functions with dimension $q$. 
The basis of the spline space, denoted by $\mbf{b}$, is chosen to be
orthonormal; that is, the elements of $\mbf{b}= \{b_1, \dots, b_q\}\trans $
satisfy $\int_I b_j(t)b_l(t)\,dt = \delta_{jl}$, or collectively,
$\int_I \mbf{b}(t)\mbf{b}(t)\trans \, dt = \mbf{I}$. 
The construction of an orthonormal basis using B-splines follows the procedure described in Appendix 1 of \cite{zhou2008joint}.
Let $\bthmu$ and $\bthnu$ be $q$-dimensional vectors of
spline coefficients such that
\begin{eqnarray}\label{eq:spline-means}
\mu(t)=\mbf{b}(t)\trans \bthmu, \quad 
\nu(t)=\mbf{b}(t)\trans \bthnu,\quad t \in I.
\end{eqnarray}
Let $\bThf$
and $\bThg$ be respectively $q\times k_\alpha$
and $q\times k_\beta$ matrices of spline coefficients
such that
\begin{eqnarray}\label{eq:spline-pc}
\mbf{f}(t)\trans  = \mbf{b}(t)\trans \bThf , \quad
\mbf{g}(t)\trans  = \mbf{b}(t)\trans \bThg, \quad t \in I.
\end{eqnarray}
In general, the basis expansions in \eqref{eq:spline-means}
and \eqref{eq:spline-pc} are only approximations. Assuming
the functions are smooth, the approximations can be very
good if a sufficiently large $q$ is used \citep{de1978practical}. 
To make the model identifiable, we require that 
$\bThf\trans \bThf = \mathbf{I}$
and $\bThg\trans \bThg=\mathbf{I}$,
then we have that 
\begin{eqnarray*}
\int_I \mbf{f}(t) \mbf{f}(t)\trans \,dt &=& \bThf \trans \int_I \mbf{b}(t)\mbf{b}(t)\trans \,dt\,\bThf = \mathbf{I}, \qquad \\
\int_I \mbf{g}(t) \mbf{g}(t)\trans \,dt &=& \bThg \trans \int_I \mbf{b}(t)\mbf{b}(t)\trans \,dt\,\bThg = \mathbf{I}.
\end{eqnarray*}

In practice, the functions are observed at discrete points (referred to as times) and the observation times can be different for different functions. Suppose there are $n$ independent pairs of functional data $(Y_i, Z_i)$, $i \in \{1, \cdots, n\}$. Let $t_{i1}, \dots, t_{in_i}$ and $s_{i1}, \dots, s_{im_i}$ denote respectively the observation times of $Y_i$ and $Z_i$, and thus 
$Y_i(t_{ij}),\ j\in\{1,\ldots,n_i\}$ and $Z_i(s_{ik}),\ k\in\{1,\ldots,m_i\}$ are the observations. Write
$\bY_i = \{Y_i(t_{i1}),\dots, Y_i(t_{in_i})\}\trans $ and $\bZ_i = \{Z_i(s_{i1}),\dots, Z_i(s_{im_i})\}\trans $.
Let $\bBy_i = \{\mbf{b}(t_{i1}), \dots, \mbf{b}(t_{in_i})\}\trans$, $\bBz_i = \{\mbf{b}(s_{i1}), \dots, \mbf{b}(s_{im_i})\}\trans$, $\bs{\epsilon_i}=\{\epsilon(t_{i1}), \dots, \epsilon(t_{in_i}) \}\trans$, $\bs{\xi_i}=\{\xi(s_{i1}), \dots, \xi(s_{im_i}) \}\trans$.
Plugging basis expansion \eqref{eq:spline-means} and \eqref{eq:spline-pc}
into model \eqref{eq:mixed-effects}, we see
that the robust reduced-rank mixed-effects model implies the following model
for observed functional data, 
\begin{eqnarray}\label{eq:bivar-smn}
\begin{aligned}
\bY_i &= \bBy_i \bthmu + \bBy_i \bThf \balpha_i + \bs{\epsilon}_i,\\
\bZ_i &= \bBz_i \bthnu + \bBz_i \bThg \bbeta_i + \bs{\xi}_i,\\
\end{aligned}
\end{eqnarray}
where $(\balpha_i\trans,\bbeta_i\trans)\trans$ are the PC scores, for $i\in\{1,\ldots,n\}$. Let $U_i$ be the latent variable corresponding to $(Y_i, Z_i)$, it follows from \eqref{eq:slash-model} and \eqref{eq:error} that, for $U_i$ iid with distribution $\mathcal{H}_\gamma$ and $i\in\{1,\ldots,n\}$,
\begin{eqnarray}\label{eq:obs_slash}
\begin{aligned}
\begin{pmatrix}\balpha_i\\ \bbeta_i\end{pmatrix}\bigg  \vert U_i = u_i &\sim \mathcal{N}\biggl(0, \frac{1}{u_i}\bSigab\biggr),  \\
  \bs{\epsilon}_i \vert U_i=u_i \sim & \mathcal{N}\biggl(0, \frac{\sigma^2_\epsilon}{u_i}  \mathbf{I}_{n_i}\biggr), \quad 
  \bs{\xi}_i \vert U_i=u_i \sim \mathcal{N}\biggl(0, \frac{\sigma^2_\xi}{u_i}  \mathbf{I}_{m_i}\biggr).
  \end{aligned}
\end{eqnarray}
%
%

\section{Model estimation}\label{sec:estimation}

\subsection{Penalized Likelihood Estimation}\label{sub:penlik}

In order to have enough flexibility of modelling arbitrary smooth
functions, the number of spline basis functions, $q$, needs to be
sufficiently large.
Following a common strategy in the nonparametric function estimation
literature, e.g., 
\cite{eilers1996flexible},
we employ roughness penalties
in a penalized likelihood formulation to prevent overfitting.

Let $L_i(\Xi;\bY_i,\bZ_i)$ denote the contribution to the likelihood
from subject $i$, where $\Xi$ denote collectively unknown parameters $\bthmu$, $\bthnu$, $\bThf$, $\bThg$, $\sigma_\epsilon^2$, $\sigma_\xi^2$, $\bDa$, $\bDb$, $\mbf{C}$, and $\gamma$ from the distribution $\mathcal{H}$.  
The likelihood for all observations is
$L(\Xi;\bY,\bZ) = L_1(\Xi;\bY_i,\bZ_i)\times\ldots\times L_n(\Xi;\bY_i,\bZ_i).$

The method of penalized log-likelihood minimizes the criterion
\begin{eqnarray}\label{eq:pen-lik}
 -\frac{2}{n} \sum_{i=1}^n\log \{L_i(\Xi;\bY_i, \bZ_i)\} 
+ {\sf PEN}.
\end{eqnarray}
We propose that the roughness penalty, PEN, take the form of integrated
squared second derivatives, i.e.,
\begin{align*}
{\sf PEN} = 
&\lambda_\mu \int_I \{\mu''(t)\}^2 \, dt
+ \lambda_f  \sum_{j=1}^{k_\alpha} \int_I \{f_j''(t)\}^2\, dt \\
& \qquad + \lambda_\nu \int_I \{\nu''(t)\}^2 \, dt
+ \lambda_g \sum_{j=1}^{k_\beta} \int_I \{g_j''(t)\}^2\, dt,
\end{align*}
where $\lambda_\mu$, $\lambda_\nu$,
$\lambda_f$, and $\lambda_g$ are four penalty parameters.
Applying the basis expansions 
\eqref{eq:spline-means} and \eqref{eq:spline-pc}, the penalty term can be written as
\begin{align}\label{eq:pen}
&{\sf PEN}(\bthmu, \bthnu, \bThf, \bThg) \notag\\
&\quad = \lambda_\mu \bthmu\trans \mbf{\Omega} \,\bthmu
+ \lambda_f \sum_{j=1}^{k_{\alpha}} \bThfj \trans  \mbf{\Omega} \,\bThfj 
+ \lambda_\nu \bthnu\trans \mbf{\Omega} \,\bthnu
+ \lambda_g \sum_{j=1}^{k_{\beta}} \bThgj \trans  \mbf{\Omega} \,\bThgj,
\end{align}
where $\mbf{\Omega} = \int_I \mbf{b}''(t) \mbf{b}''(t)\trans \,dt$ is the penalty matrix, 
and $\bThfj$ and $\bThgj$ are, respectively, the $j$-th columns of
$\bThf$ and $\bThg$. Our use of four penalty parameters gives the 
flexibility of allowing different amounts of smoothing for the mean
functions and PC functions.

Direct minimization of the penalized log-likelihood \eqref{eq:pen-lik} is challenging due to the complication of the likelihood evaluation. We use the latent variable representation of the model given in \eqref{eq:bivar-smn} and  \eqref{eq:obs_slash}  and then apply the EM algorithm \citep{dempster1977maximum} to minimize the objective function \eqref{eq:pen-lik}. Treating the PC scores $\balpha_i,\ \bbeta_i$ and the latent scale variables $u_i$ as missing data, we obtain the following complete data likelihood function
\begin{align*}
L_i(\Xi;\bY_i,\bZ_i,\balpha_i,\bbeta_i,u_i)
& = p_y(\bY_i|\balpha_i,u_i;\bthmu,\bThf,\sigma_\epsilon^2 ) p_z(\bZ_i|\bbeta_i,u_i;\bthnu,\bThg,\sigma_\xi^2)  \\
& \quad \times p_s(\bbeta_i,\balpha_i|u_i;\bDa, \bDb, \mbf{C}) p_u(u_i;\gamma),	
\end{align*}
where $p_y,p_z,p_s$ are the conditional density functions of $\bY_i,\ \bZ_i,$ and $(\balpha_i\trans,\bbeta_i\trans)$, respectively;  $p_u$ is the density function of the latent variable $u_i$.
With an irrelevant
constant ignored, it follows that
\begin{eqnarray}\label{eq:comp-log-lik}
\begin{aligned}
 -2 \, & \log\{L_i(\Xi;\bY_i,\bZ_i,\balpha_i,\bbeta_i,u_i)\}\\
& = n_i \log \,\sigma_\epsilon^2  - n_i \log\, u_i  + m_i \log\,\sigma_\xi^2 - m_i \log\,u_i\\
& \qquad +\frac{u_i}{\sigma_\epsilon^2}
(\bY_i-\bBy_i \bthmu- \bBy_i \bThf \balpha_i)\trans
(\bY_i-\bBy_i \bthmu- \bBy_i \bThf \balpha_i) \\
& \qquad + \frac{u_i}{\sigma_\xi^2}
(\bZ_i - \bBz_i \bthnu - \bBz_i \bThg \bbeta_i)\trans
(\bZ_i - \bBz_i \bthnu - \bBz_i \bThg \bbeta_i)\\
& \qquad + \log\,|\bSigab| - (k_\alpha+k_\beta)
\log\, u_i + u_i\, (\balpha_i\trans\,\bbeta_i\trans) \bSigab^{-1}
 \begin{pmatrix}\balpha_i\\ \bbeta_i\end{pmatrix}-2\log p_u(u_i;\gamma).
\end{aligned}
\end{eqnarray}

Given the current estimate of the parameter set at step $\ell$, denoted as $\Xi^{(\ell)}$,  
the EM algorithm minimizes with respect to the parameter set $\Xi$ the following penalized conditional expected log-likelihood
\begin{eqnarray}\label{eq:pen-expec-lik}
\begin{aligned}
 &-\frac{2}{n} \,\sum_{i=1}^n E_{(\balpha_i,\bbeta_i,U   _i)}[\log\{L_i(\Xi;\bY_i,\bZ_i,\balpha_i,\bbeta_i,u_i)\}|\bY_i, \bZ_i;
\Xi^{(\ell)}]\\
&\qquad + {\sf PEN}(\bthmu, \bthnu, \bThf, \bThg),
\end{aligned}
\end{eqnarray}
to obtain an update $\Xi^{(\ell+1)}$. The algorithm iterates until convergence is reached. 
Under some regularity conditions, the algorithm is guaranteed to
converge to a local minimizer of the penalized log-likelihood
\eqref{eq:pen-lik} \citep{wu1983convergence}. Details of the EM
algorithm are given in Appendix~A.

\subsection{Model Selection}\label{sec:tuning}

\textit{Specification of B-Splines.}
The number of knots and the positions of the knots are not crucial in
many applications as long as sufficiently many knots are used to cover
the data range, since the roughness penalty helps regularizing the
estimation and prevent overfitting 
\citep{eilers1996flexible}.
We found that using 10-20 knots is often sufficient.

\textit{Choice of Penalty Parameters.}
When the numbers of PCs are fixed, $K$-fold within-subject cross-validation (CV) is used to select the penalty parameters, using the mean absolute error as the metric to
measure goodness-of-fit.
The downhill simplex method of \cite{nelder1965}
is used to search the optimal penalty parameters.

\textit{Numbers of PCs.}
Within-subject CV can also be used to select the
numbers of PCs. Specifically, for each pair of $(k_\alpha, k_\beta)$, let 
$\hat{\lambda}_\mu, \hat{\lambda}_\nu, \hat{\lambda}_f,
\hat{\lambda}_g$ be the selected penalty parameters that minimize the
within-subject CV value, and 
denote the corresponding CV value as
$CV(k_\alpha,k_\beta;\hat{\lambda}_\mu, \hat{\lambda}_\nu,
\hat{\lambda}_f, \hat{\lambda}_g)$. We may choose the numbers of PCs as
$$(\widehat{k}_\alpha, \widehat{k}_\beta) = \arg\min_{k_\alpha,
  k_\beta} \mathrm{CV} (k_\alpha,k_\beta;\hat{\lambda}_\mu,\ \hat{\lambda}_\nu, \hat{\lambda}_f, \hat{\lambda}_g).$$
Our experience suggests that $(\widehat{k}_\alpha, \widehat{k}_\beta)$
may be larger than necessary. Inspired by the idea of Cattell's scree test \citep{cattell1966scree},
we propose to choose the smallest $k_\alpha, k_\beta$ whose CV value
is not greater than the smallest CV value by a factor of $r$. To be specific,
denote the set 
$$S(r)={\{(k_\alpha,k_\beta): CV( k_\alpha,k_\beta; \ \cdot) \le (1+r)CV( \widehat{k}_\alpha, \widehat{k}_\beta;\ \cdot)\}}.$$
The numbers of PCs $k_\alpha, k_\beta$ can be chosen as
$$(\widetilde{k}_\alpha,\widetilde{k}_\beta) = \arg \min_{(k_\alpha,k_\beta)\in S(r)} (k_\alpha+k_\beta).$$

In our simulation study, we found that including a small $r>0$ as above
substantially improves the ability of CV to choose the correct numbers of significant PCs.  See
Table~\ref{tb:correct_rate} below and its discussion.

\section{Simulation}\label{sec:simu}
In Section \ref{sub:setup}, we present details of the simulation setups. The performance of model fitting and the selection of the numbers of PCs are shown in Section \ref{sub:result}.

\subsection{Simulation Setup}\label{sub:setup}
In this simulation study, we generated the simulation data based on equations \eqref{eq:mixed-effects},
\eqref{eq:slash-model} and \eqref{eq:error} with $t\in [0,1]$ and two PCs, $k_\alpha=k_\beta=2$.
The mean functions are given, for all $t \in [0,1]$, by
\begin{align*}
\mu(t) & = 2.5+ 2.5t+ 2.5\,\exp\{-20(t-0.6)^2\},\\
\nu(t) & = 7.5 - 1.5t - 2.5\,\exp\{-20(t-0.3)^2\}.
\end{align*}
The PC functions are given, for all $t \in [0,1]$, by
\begin{align*}
f_{1}(t) & =\frac{\sqrt{15}}{1+\sqrt{5}}(t^2+\frac{1}{\sqrt{5}}),  \quad  
f_{2}(t)=\frac{\sqrt{15}}{\sqrt{5}-1}(t^2-\frac{1}{\sqrt{5}}),\\
g_{1}(t) & =\sqrt{2} \cos(2\pi t), \quad \qquad\ \ 
g_{2}(t)=\sqrt{2} \sin(2 \pi t),
\end{align*} 
and
\begin{eqnarray*}
\bSigab = 
\begin{pmatrix}
1 & 0 & -0.4 & 0.12 \\
0 & 0.25 & 0.15 & -0.05 \\
-0.4 & 0.15 & 1.44 & 0 \\
0.12 & -0.05 & 0 & 0.36
\end{pmatrix}.
\end{eqnarray*}
We considered  two levels of the scale parameters: $\sigma_\epsilon^2= \sigma^2_\xi = 0.04$, $\sigma_\epsilon^2= \sigma^2_\xi = 0.25$.

We considered 3 scenarios of generating data from the models given in Section~\ref{sec:model}.

\begin{itemize}
    \item Scenario 1: The latent variable $u\equiv 1$. Note that this model is an RRME-normal model.
	\item Scenario 2: The latent variable $u\sim \text{Gamma}(\gamma/2,\gamma/2)$, where $\gamma$ is the degrees of freedom. Note that this model is an RRME-t model. We considered three different values of $\gamma$, $\gamma=2,\ 5$ and $10$. 
	\item Scenario 3: The latent variable $u\sim \text{Beta}(\gamma,1)$, where $\gamma$ is the degrees of freedom.  Note that this model is an RRME-slash model. We considered three different values of $\gamma$, $\gamma=1,\ 2$ and $5$.
\end{itemize} 

We also considered 2 scenarios of generating data by adding outliers to data generated from the RRME-normal model. 
In one scenario, the outliers of functional data are outlying measurements at some points, and in another scenario,  they are outlying shapes. We used these two scenarios to evaluate the performance of the proposed method when the data were not generated from the model that the method is designed for.

\begin{itemize}
	\item Scenario 4: First, we generated data from an RRME-normal model.  Next, we randomly selected 5\% of the observations from $Y_i(t_{ij})$ and $Z_i(t_{ij}), i\in\{1,\ldots,n\}, j\in\{1,\ldots,n_i\}$, then with probability $1/2$ added or subtracted a random number generated from Uniform$(8, 10)$ to each of the selected observations. The selected observations are outlying measurements.   
    \item Scenario 5:  First, we generated data from an RRME-normal model.   
Next, we randomly selected 5\% pairs of functions from $(Y_i, Z_i)$, $i\in\{1,\ldots,n\}$, and to each of the PC scores of the selected functions added a random number independently generated from Uniform$(-4,4)$. 
          The selected functions have outlying shapes.  
\end{itemize}

\begin{figure}
\centering
\includegraphics[width = 0.85\textwidth]{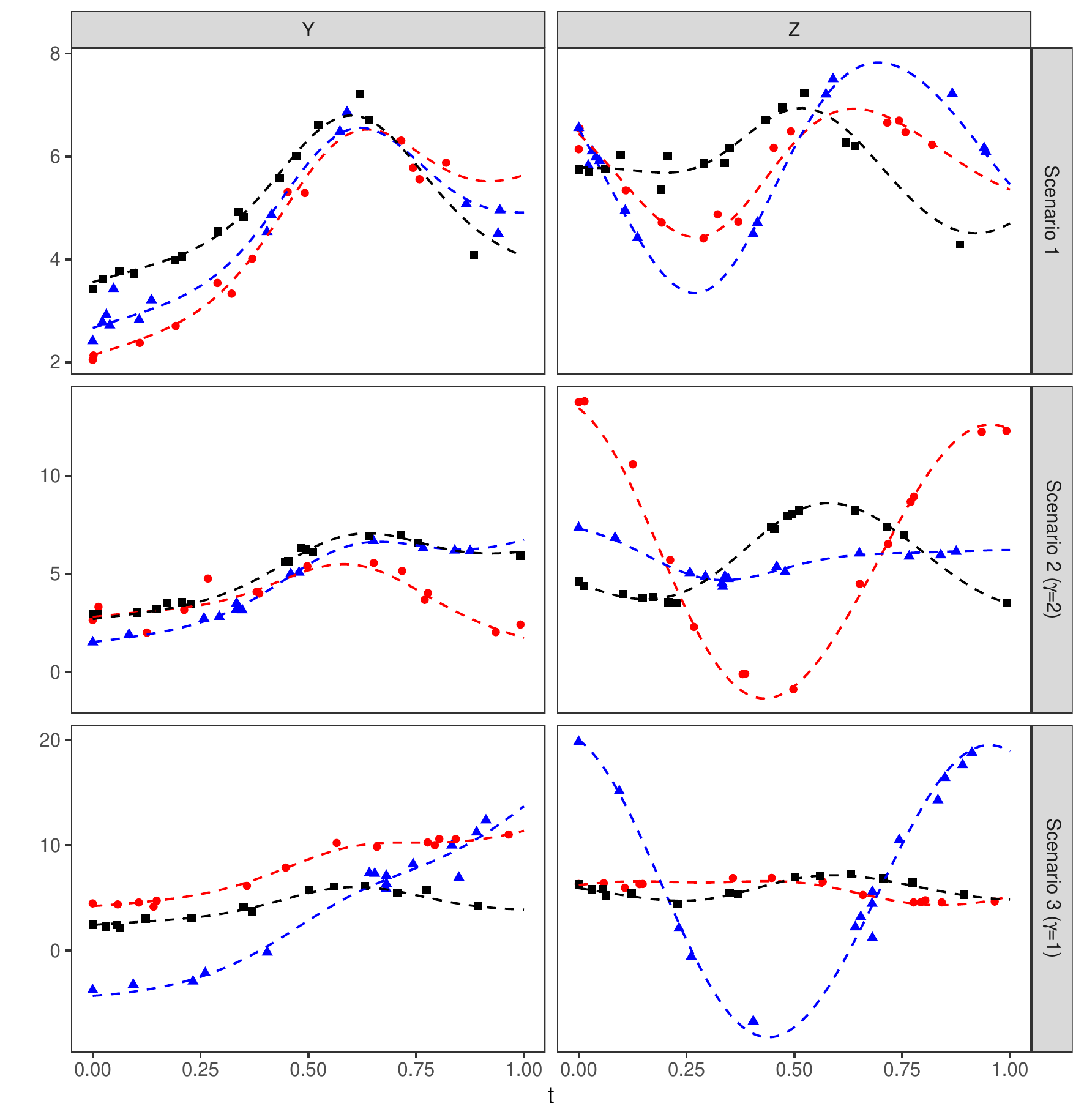}
\caption{Observations and unobserved true functions from sparsely observed paired functional data generated from Scenarios 1-3, for $\sigma_\epsilon^2= \sigma^2_\xi =0.04$. Each row, corresponding to one scenario, shows data of three selected pairs of functions indicated using different colors and dot types. The unobserved true functions are shown as dashed lines, the noisy observations are shown as dots.}
\label{fig:simu_data}
\end{figure}

\begin{figure}
\centering
\includegraphics[width=0.85\textwidth]{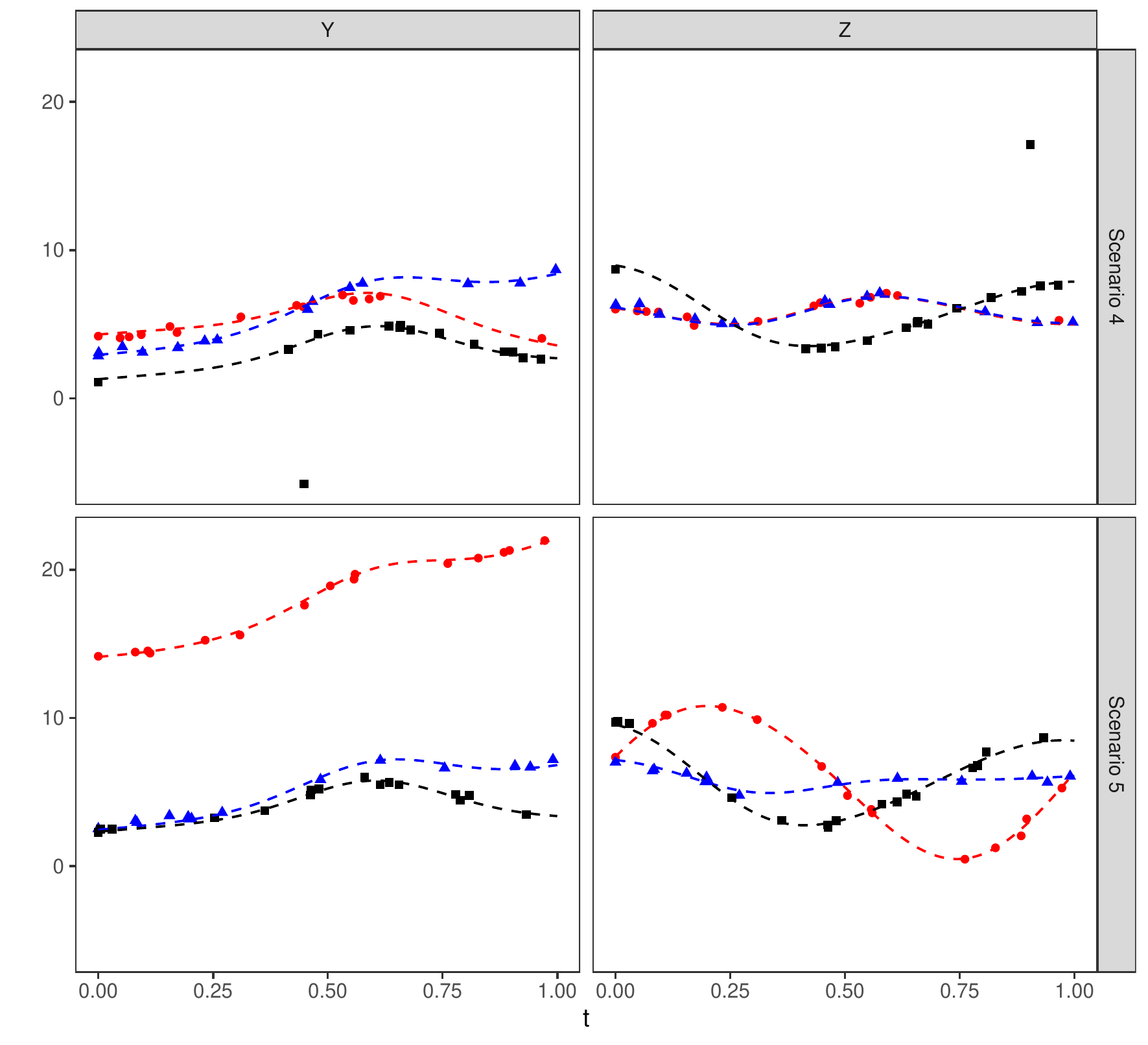}
\caption{
Observations and unobserved true functions from sparsely observed paired functional data generated from Scenarios 4-5, for $\sigma_\epsilon^2= \sigma^2_\xi =0.04$. Each row, corresponding to one scenario, shows data of three selected pairs of functions indicated using different colors and dot types. The unobserved true functions are shown as dashed lines, the noisy observations are shown as dots.
}
\label{fig:outlier}
\end{figure}

In each simulation setup, we generated data for $n=100$ subjects.
For each subject, the number of observed time points was randomly generated from $1+ \text{Binomial}(15, 0.9)$; 
the first time point was set at $0$ and the rest were generated independently from Uniform$(0,1)$. Figure~\ref{fig:simu_data} shows observations from three selected pairs of functions in Scenarios 1-3. Figure~\ref{fig:outlier} shows observations from three selected pairs of functions in Scenarios 4 and 5. 

\subsection{Simulation results}\label{sub:result}
To evaluate the performance of our proposed method, we ran the simulation $m=500$ times for each setup. We applied our method corresponding to the RRME-t, RRME-slash and RRME-normal models to each simulated dataset. 
We used cubic B-splines with 10 interior knots for the basis
functions. The penalty parameters were selected using a 10-fold within-subject CV and the downhill simplex method for optimization. 

To quantitatively measure the difference between a function $h$ and its estimator $\hat{h}$ on the interval $[0,1]$,
we used the integrated absolute error (IAE) defined as 
\begin{eqnarray*}
  {\rm IAE} = \int_0^1 |\hat{h}(t)-h(t)| dt.
\end{eqnarray*}
In our implementation, the integral is calculated using the Riemann sum where the interval of integration is partitioned into $100$ equal-length intervals. IAE is directly used for evaluating the estimation of the mean function and the PC functions. When evaluating the estimation of individual (or subject level) functions, we take the average of IAEs over all individual functions.

We first applied the proposed method using the true numbers of PCs, $(k_\alpha,k_\beta)=(2,2)$.
Table~\ref{tab:IMSE_set1} and Table~\ref{tab:IMSE_set1-2} present the average IAEs for the mean functions, the PC functions and the individual functions, under five simulation scenarios. 
In general, the average IAEs are smaller for $\sigma_\epsilon^2=\sigma_\xi^2= 0.04$, when compared with the corresponding setups for $\sigma_\epsilon^2=\sigma_\xi^2= 0.25$.
The RRME-t models and RRME-slash models performed similarly.
When the data were generated from Scenario 1, the average IAEs using the RRME-t, the RRME-slash and the RRME-normal model are similar.
When the data were generated from Scenarios 2 and 3, compared with using the RRME-normal model, using the RRME-t and RRME-slash model gave similar average IAEs on the mean functions, but smaller average IAEs on the PC functions and individual functions. In particular,
when the data were generated from the Scenario 2 with $\gamma=2$ or Scenario 3 with $\gamma=1$, the average IAEs using the RRME-t or the RRME-slash model are smaller than those using the RRME-normal model by 62\%-69\% on PC functions, and by 12\%-22\% on individual functions. However, as the degrees of freedom increases, the average IAEs using the RRME-t or RRME-slash model get closer to that using the RRME-normal model. 
When the data were generated from Scenario 4, compared with average IAEs using the RRME-normal model, the average IAEs using the RRME-t or RRME-slash model are smaller by 15\%-30\% on mean functions, by 9\%-65\% on PC functions, and by 6\%-11\% on  individual functions. 
When the data were generated from Scenario 5, compared with average IAEs using RRME-normal model, the average IAEs using RRME-t or RRME-slash model are smaller by 14\%-29\% on PC functions, and similar on mean and individual functions.

From Table~\ref{tab:IMSE_set1} and Table~\ref{tab:IMSE_set1-2}, we observe that the performance of the proposed robust method is not sensitive to the choice of the Student $t$ or the slash distribution. This phenomenon is also observed in our extensive simulation studies not reported here. In practice, one can use either the t or the slash distribution. We would recommend using the Student $t$ distribution because it is more familiar to statisticians and the calculation is simpler.

Next we evaluate the proposed method on the selection of the number of PCs using datasets simulated from Scenario 2 with degrees of freedom $\gamma=2,5$ and Scenario 3 with $\gamma=1,2$, with two levels of scale parameters.
Data simulated from Scenario 2 were fitted with RRME-t model and data simulated from Scenario 3 were fitted with RRME-slash model. For all model fitting, 
we used cubic B-splines with 10 interior knots for the basis functions. The penalty parameters were selected using 10-fold CV and the downhill simplex method. The numbers of PCs were selected from $\{(k_\alpha,k_\beta):1\le k_\alpha\le 3,\ 1\le k_\beta\le 3\}$ with the parameter $r=0$, $0.01$, and $0.05$. 
From Table~\ref{tb:correct_rate}, we see that in both scenarios, when $r>0$, the CV does a good job in selecting the correct 
number of significant PCs and the result is not sensitive to the choice of $r$, but the CV does not work well when $r=0$. We recommend using $r=0.01$ or $0.05$ in practice.

\begin{table}[!ht]
\renewcommand\arraystretch{0.5}    
\begin{center}
\caption{
Results for simulation Scenarios 1-3. Reported are the average integrated absolute errors (IAEs), based on $500$ simulation runs, for estimating the mean functions,
the PC functions, and the individual functions using the RRME-t, RRME-slash and RRME-normal models. Each reported number equals the actual number multiplied by $1000$.} 
\label{tab:IMSE_set1} \vspace{10pt}
\scalebox{0.8}{
\begin{tabular}{*{12}{c}}
\hline\hline\\[-0.7em]
\multicolumn{3}{c}{Simulation setup} & Fitting model &\multicolumn{2}{c}{Mean functions} & 
\multicolumn{2}{c}{PC of Y} & \multicolumn{2}{c}{PC of Z} & \multicolumn{2}{c}{Individual functions}\\
 & $\sigma_\epsilon^2=\sigma_\xi^2$ & $\gamma$ & & $\mu$ & $\nu$ &$f_1$ &$f_2$ &$g_1$ &$g_2$ & $Y$ & $Z$ \\[0.3em]
\hline\\[-0.7em] 
Scenario 1 & 0.04 &  & RRME-t &  87.72 & 108.51 &  51.34 &  55.01 &  49.81 &  52.42 &  63.56 &  65.48 \\ 
 &  &  & RRME-slash &  87.70 & 108.52 &  51.14 &  54.79 &  49.77 &  52.36 &  63.56 &  65.47 \\ 
 &  &  & RRME-normal &  87.70 & 108.52 &  51.09 &  54.79 &  49.77 &  52.37 &  63.57 &  65.47 \\ 
\hline\\[-0.7em] 
 & 0.25 &  & RRME-t &  92.75 & 112.41 &  57.31 &  66.65 &  57.58 &  65.45 & 153.60 & 158.62 \\ 
 &  &  & RRME-slash &  92.76 & 112.40 &  57.02 &  66.30 &  57.38 &  65.33 & 153.58 & 158.62 \\ 
 &  &  & RRME-normal &  92.71 & 112.39 &  57.01 &  66.38 &  57.41 &  65.42 & 153.60 & 158.61 \\ 
 \hline \hline\\[-0.7em]
Scenario 2 & 0.04 &  2 & RRME-t & 244.84 & 289.26 &  61.03 &  65.30 &  62.90 &  65.40 & 112.26 & 114.92 \\ 
 &  &  & RRME-slash & 244.84 & 289.26 &  60.69 &  65.09 &  62.17 &  64.72 & 112.27 & 115.00 \\ 
 &  &  & RRME-normal & 246.89 & 289.93 & 172.52 & 180.73 & 191.91 & 198.59 & 131.94 & 144.91 \\ 
 &  &  5 & RRME-t &  43.41 &  53.81 &  52.29 &  55.90 &  54.94 &  58.01 &  30.34 &  31.27 \\ 
 &  &  & RRME-slash &  43.42 &  53.82 &  52.21 &  55.79 &  54.96 &  58.06 &  30.35 &  31.29 \\ 
 &  &  & RRME-normal &  43.72 &  54.20 &  72.63 &  76.75 &  76.10 & 118.94 &  31.16 &  41.88 \\ 
 &  & 10 & RRME-t &  20.67 &  24.01 &  51.48 &  54.94 &  52.25 &  55.07 &  13.95 &  14.36 \\ 
 &  &  & RRME-slash &  20.67 &  24.02 &  51.77 &  55.35 &  52.76 &  55.53 &  13.97 &  14.38 \\ 
 &  &  & RRME-normal &  20.73 &  24.16 &  54.91 &  59.14 &  57.22 & 107.03 &  14.19 &  20.45 \\ 
 \hline\\[-0.7em] 
 & 0.25 &  2 & RRME-t & 246.96 & 291.22 &  67.18 &  78.09 &  70.84 &  78.06 & 272.07 & 279.32 \\ 
 &  &  & RRME-slash & 246.83 & 291.12 &  66.87 &  77.98 &  70.28 &  77.47 & 272.24 & 279.32 \\ 
 &  &  & RRME-normal & 260.62 & 297.00 & 189.46 & 206.04 & 213.68 & 248.15 & 324.20 & 357.55 \\ 
 &  &  5 & RRME-t &  45.27 &  55.03 &  57.25 &  66.49 &  62.10 &  70.93 &  73.48 &  75.80 \\ 
 &  &  & RRME-slash &  45.30 &  55.07 &  57.50 &  66.57 &  61.98 &  70.91 &  73.55 &  75.85 \\ 
 &  &  & RRME-normal &  46.68 &  56.41 &  78.77 &  89.57 &  85.02 & 200.00 &  75.27 & 100.81 \\ 
 &  & 10 & RRME-t &  21.98 &  24.99 &  57.23 &  65.56 &  59.73 &  68.39 &  33.71 &  34.82 \\ 
 &  &  & RRME-slash &  22.00 &  25.00 &  57.53 &  65.82 &  59.95 &  68.55 &  33.74 &  34.83 \\ 
 &  &  & RRME-normal &  22.27 &  25.31 &  62.43 &  74.30 &  65.06 & 130.98 &  34.59 &  41.17 \\
\hline\hline\\[-0.7em] 
Scenario 3 & 0.04 &  1 & RRME-t & 268.43 & 314.68 &  56.48 &  60.12 &  60.18 &  63.02 & 127.07 & 131.40 \\ 
 &  &  & RRME-slash & 268.56 & 314.64 &  56.99 &  60.75 &  59.84 &  62.64 & 130.54 & 131.44 \\ 
 &  &  & RRME-normal & 271.93 & 314.72 & 160.49 & 166.51 & 161.35 & 168.08 & 151.90 & 163.94 \\ 
 &  &  2 & RRME-t & 116.36 & 145.79 &  52.18 &  55.57 &  51.47 &  53.98 &  85.14 &  87.84 \\ 
 &  &  & RRME-slash & 116.35 & 145.77 &  52.36 &  55.71 &  51.89 &  54.40 &  85.15 &  87.81 \\ 
 &  &  & RRME-normal & 116.72 & 146.04 &  64.99 &  69.25 &  67.10 &  70.09 &  86.35 &  90.13 \\ 
 &  &  5 & RRME-t & 102.17 & 124.04 &  52.99 &  56.84 &  50.66 &  53.44 &  71.00 &  73.02 \\ 
 &  &  & RRME-slash & 102.18 & 124.04 &  53.13 &  56.98 &  50.87 &  53.68 &  71.01 &  73.03 \\ 
 &  &  & RRME-normal & 102.23 & 124.04 &  53.64 &  57.38 &  51.66 &  54.42 &  71.03 &  73.12 \\ 
 \hline\\[-0.7em] 
 & 0.25 &  1 & RRME-t & 272.24 & 314.32 &  61.49 &  71.02 &  66.80 &  75.06 & 307.20 & 318.98 \\ 
 &  &  & RRME-slash & 272.05 & 314.24 &  61.61 &  71.06 &  66.51 &  74.65 & 307.12 & 318.86 \\ 
 &  &  & RRME-normal & 293.01 & 317.84 & 168.31 & 187.22 & 180.47 & 211.31 & 381.95 & 408.37 \\ 
 &  &  2 & RRME-t & 122.11 & 149.76 &  57.72 &  66.01 &  58.70 &  67.34 & 205.94 & 213.60 \\ 
 &  &  & RRME-slash & 122.08 & 149.63 &  57.79 &  66.20 &  59.09 &  66.59 & 205.87 & 212.79 \\ 
 &  &  & RRME-normal & 123.99 & 151.05 &  72.01 &  81.88 &  76.09 &  85.33 & 208.61 & 217.93 \\ 
 &  &  5 & RRME-t & 107.95 & 127.97 &  58.40 &  66.95 &  58.14 &  66.55 & 171.74 & 176.94 \\ 
 &  &  & RRME-slash & 107.96 & 127.95 &  58.53 &  67.00 &  58.24 &  66.71 & 171.71 & 176.91 \\ 
 &  &  & RRME-normal & 108.10 & 127.95 &  59.28 &  67.91 &  59.37 &  67.72 & 171.88 & 177.14 \\ 
\hline\hline\\[-0.7em] 
\end{tabular}}
\end{center}
\end{table}

\begin{table}[!ht]
\renewcommand\arraystretch{0.5}    
\begin{center}
\caption{
Results for simulation Scenarios 4 and 5. Reported are the average integrated absolute errors (IAEs), based on $500$ simulation runs, for estimating the mean functions,
the PC functions, and the individual functions using the RRME-t, RRME-slash and RRME-normal models. Each reported number equals the actual number multiplied by $1000$.} 
\label{tab:IMSE_set1-2} \vspace{10pt}
\scalebox{0.8}{
\begin{tabular}{*{12}{c}}
\hline\hline\\[-0.7em]
\multicolumn{2}{c}{Simulation setup} & Fitting model &\multicolumn{2}{c}{Mean functions} & 
\multicolumn{2}{c}{PC of Y} & \multicolumn{2}{c}{PC of Z} & \multicolumn{2}{c}{Individual functions}\\
 & $\sigma_\epsilon^2=\sigma_\xi^2$ & & $\mu$ & $\nu$ &$f_1$ &$f_2$ &$g_1$ &$g_2$ & $Y$ & $Z$ \\[0.3em]
\hline\\[-0.7em]
Scenario 4 & 0.04 &  RRME-t & 106.62 & 128.40 &  99.17 & 114.69 & 110.72 & 144.22 & 450.66 & 490.70 \\ 
 &  &  RRME-slash & 106.17 & 127.03 & 100.56 & 114.94 & 110.96 & 127.99 & 452.18 & 488.28 \\ 
 &  &  RRME-normal & 151.55 & 160.23 & 110.46 & 173.34 & 122.73 & 366.80 & 486.97 & 549.54 \\ 
 & 0.25 &  RRME-t & 120.51 & 137.60 &  91.56 & 122.55 & 104.00 & 171.68 & 465.54 & 509.92 \\ 
 &  &  RRME-slash & 120.54 & 137.58 &  91.68 & 122.29 & 103.02 & 157.01 & 466.60 & 507.66 \\ 
 &  &  RRME-normal & 154.29 & 162.40 & 111.79 & 180.13 & 125.78 & 382.84 & 498.73 & 563.67 \\
\hline\\[-0.7em] 
Scenario 5 & 0.04 &  RRME-t & 108.46 & 126.84 & 111.34 & 115.46 &  89.22 &  91.29 &  63.85 &  66.03 \\ 
 &  &  RRME-slash & 108.44 & 126.81 &  95.88 &  99.57 &  79.76 &  81.73 &  63.92 &  66.12 \\ 
 &  &  RRME-normal & 108.48 & 126.83 & 134.74 & 139.32 & 104.72 & 106.98 &  63.81 &  65.99 \\ 
 & 0.25 &  RRME-t & 112.93 & 129.97 & 111.97 & 117.43 &  92.83 &  97.33 & 155.58 & 160.84 \\ 
 &  &  RRME-slash & 112.83 & 129.88 & 100.09 & 105.57 &  85.40 &  90.04 & 155.70 & 160.95 \\ 
 &  &  RRME-normal & 113.06 & 129.98 & 136.54 & 142.48 & 108.81 & 113.08 & 155.45 & 160.73 \\ 
\hline\hline\\[-0.7em] 
\end{tabular}}
\end{center}
\end{table}

\begin{table}[!ht]
\begin{center}
\caption{Proportions of the correct choice of the numbers of PCs, based on 500
  simulation runs, for simulation Scenarios 2 \& 3.}
\renewcommand\arraystretch{0.5}  
\label{tb:correct_rate}
\begin{tabular}{cccccc}
\hline\hline\\[-0.7em]
\multicolumn{3}{c}{Simulation setup} & \multicolumn{3}{c}{r} \\\\[-0.7em]
 & $\sigma_\epsilon^2=\sigma_\xi^2$ &$\gamma$ &0&0.01&0.05\\
\hline\\[-0.7em]
Scenarios 2 &0.04&2 & 28.4\% & 99.6\% & 100.0\% \\
&&5 & 49.2\% & 100.0\% & 100.0\% \\
&0.25&2 & 21.8\% & 99.4\% & 100.0\% \\
&&5 & 39.6\% & 100.0\% & 100.0\% \\
\hline\\[-0.7em]
Scenarios 3 &0.04&1 & 27.8\% & 99.4\% & 99.8\% \\
&&2 & 41.2\% & 100.0\% & 100.0\% \\
&0.25&1 & 27.2\% & 99.6\% & 99.8\% \\
&&2 & 35.8\% & 100.0\% & 100.0\% \\
\hline\hline\\[-0.7em]
\end{tabular}
\end{center}
\end{table}

\section{Type Ia Supernova Light Curve Example}\label{sec:AIDS}
In this section, we apply the proposed method to a dataset
studied by \cite{he2018characterization}; more details are given in Supplementary Material, Section A.
This dataset contains measurements of multi-band light curves from $102$ SNeIa. 
Among which, there are 84 SNeIa normal SNeIa, 9 SN1991bg-like SNeIa,
3 super-Chandrasekhar SNeIa or "SNe Iax" whose photometric characteristics are similar to that of SN1991bg-like SNeIa, 5 overluminous subtypes with photometric features different from that of normal or SN1991bg-like SNeIa, and 1 unlabeled Type Ia supernova.

Each Type Ia supernova has multiple light curves corresponding to different astronomical filters. Essential astronomical corrections were performed and all light curves were aligned by setting the peak magnitude to zero and the Julian date of maxima to zero. The redshift ($z$) effect was removed by dividing Julian dates by a factor of $(1+z)$, transforming Julian date to phase. Therefore each light curve becomes a function of phase. We focus on analyzing the R and I band data with phases between $-10$ and $50$ in this study. 

The proposed method with RRME-t and RRME-normal models were applied on the observed R band and I band data. For the basis functions, we used cubic B-splines with $15$ knots equally placed on $[-5,45]$ and two extra knots, $-10, -7.5$ and $47.5,\ 50$, on each side of this interval; same as that in \cite{he2018characterization}.
The penalty parameters were selected by the downhill simplex method using 10-fold within-subject cross-validation (CV) with the mean absolute error (MAE) as the metric.  
Table~\ref{tb:TSE-CV} presents the CV values using different numbers of PCs.
Following the selection procedure introduced in Section \ref{sec:tuning} using $r=0.05$, with both the RRME-t models and the RRME-normal models, we selected 2 PCs for fitting the R band data and 3 PCs for fitting the I band data. 

\begin{table}
\renewcommand\arraystretch{0.5} 
\begin{center}
\caption{Ten-fold CV values of Type Ia supernova light curves data fitted with RRME-t and RRME-normal models using different numbers of PCs. Each reported CV value equals the actual number multiplied by $1000$.
}
\scalebox{0.94}{
\begin{tabular}{ cc }   
\label{tb:TSE-CV}
\begin{tabular}{ccccc}
\multicolumn{5}{c}{RRME-t model}\\ 
\hline\hline\\[-0.7em]
 & $k_\alpha=1$ & $k_\alpha=2$ & $k_\alpha=3$ & $k_\alpha=4$\\  
\hline\\[-0.7em]
$k_\beta=1$ & 161.26 & 150.82 & 145.55 & 170.06 \\ 
$k_\beta=2$ & 140.55 & 128.06 & 124.11 & 129.68 \\ 
$k_\beta=3$ & 120.15 & 106.46 & 144.27 & 126.83 \\ 
$k_\beta=4$ & 159.94 & 146.87 & 130.19 & 128.35 \\ 
\hline\hline\\[-0.7em]
\end{tabular} &

\begin{tabular}{ccccc}
\multicolumn{5}{c}{RRME-normal model}\\ 
\hline\hline\\[-0.7em]
 & $k_\alpha=1$ & $k_\alpha=2$ & $k_\alpha=3$ & $k_\alpha=4$\\ 
\hline\\[-0.7em]
$k_\beta=1$ & 161.80 & 153.20 & 172.05 & 152.14 \\ 
$k_\beta=2$ & 145.13 & 142.86 & 156.35 &  174.86\\ 
$k_\beta=3$ & 130.94 & 130.00 & 124.53 & Na \\ 
$k_\beta=4$ & 166.25 & 165.35 & 154.94 & 180.73 \\ 
\hline\hline\\[-0.7em]
\end{tabular}\\
\end{tabular}}
\end{center}
\end{table}

\begin{figure}[hbt]
\includegraphics[width=0.9\textwidth]{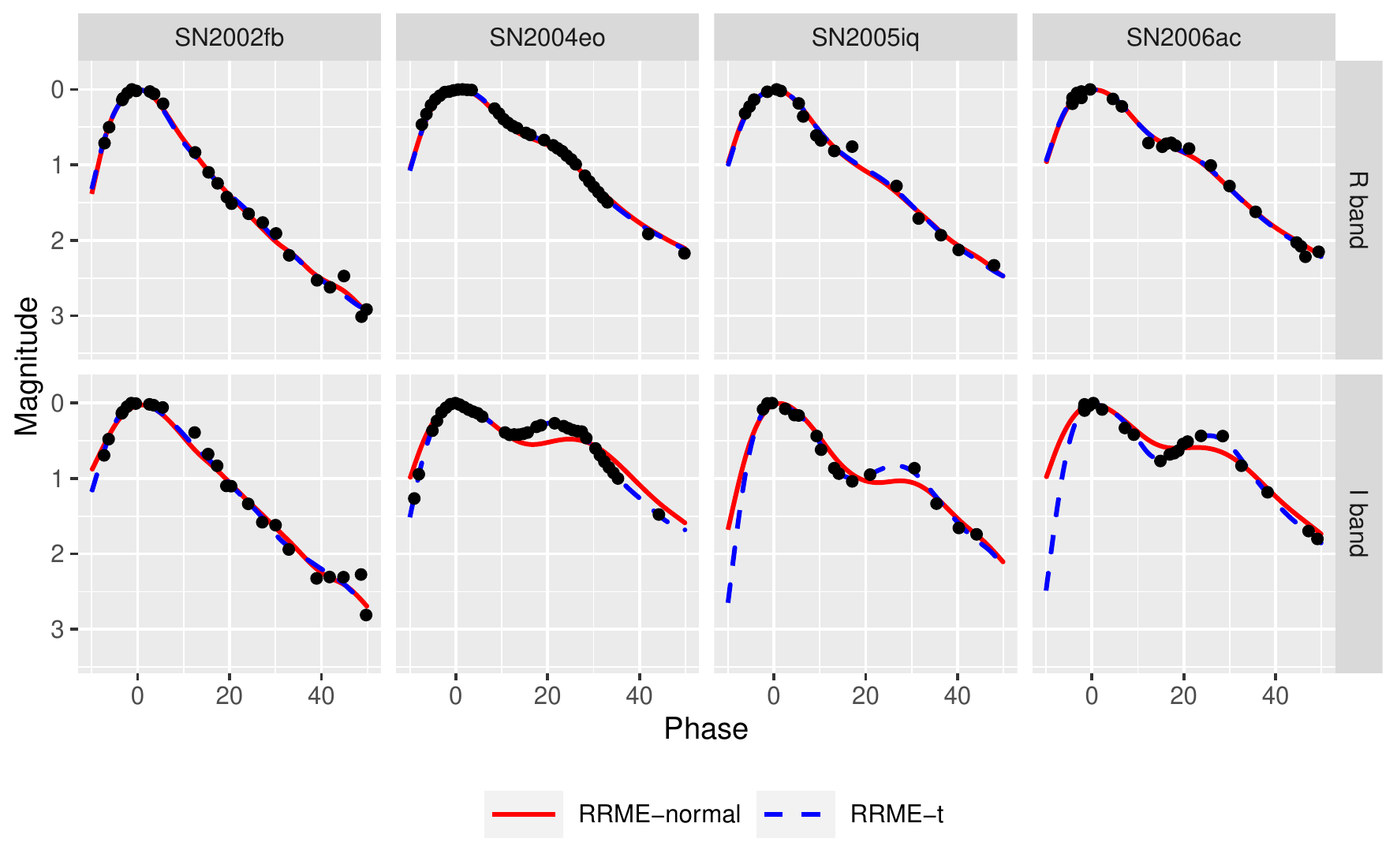}
\caption{Observations and fitted functions of one SN1991bg-like supernova, SN2002fb and three normal SNeIa: SN2004eo, SN2005iq, and SN2006ac. Black dots are observations; 
  red solid lines are fitted functions using an RRME-normal model, blue
  dashed lines are fitted functions using an RRME-t model.}
\label{fig:SNeIa}
\end{figure}

Figure~\ref{fig:SNeIa} shows the observed data and fitted light curves using the RRME-t and the RRME-normal model for one SN1991bg-like supernova SN2002fb and three normal SNeIa: SN2004eo, SN2005iq, SN2006ac.
The observed data of SN2002fb illustrate the characteristics of
SN1991bg-like SNeIa, namely the absence of a right shoulder in the R band light curves and
the missing of the secondary maxima in the I band light curves which usually
occurs around phase 20 to 35 for normal SNeIa. Both
RRME-t and RRME-normal model provide similar fits on the observed R-band data. However, the RRME-t model provides better fits for the I band data of three normal SNeIa, whose secondary peak features are not fully captured by the RRME-normal model. 
Ten-fold within-subject CV value using the RRME-t model is $0.48$ for the R band data
and $0.58$ for the I band data, while using the RRME-normal model it is $0.50$ for the 
R band data and $0.80$ for the I band data. 

\begin{figure}[hbt]
\includegraphics[width=\textwidth]{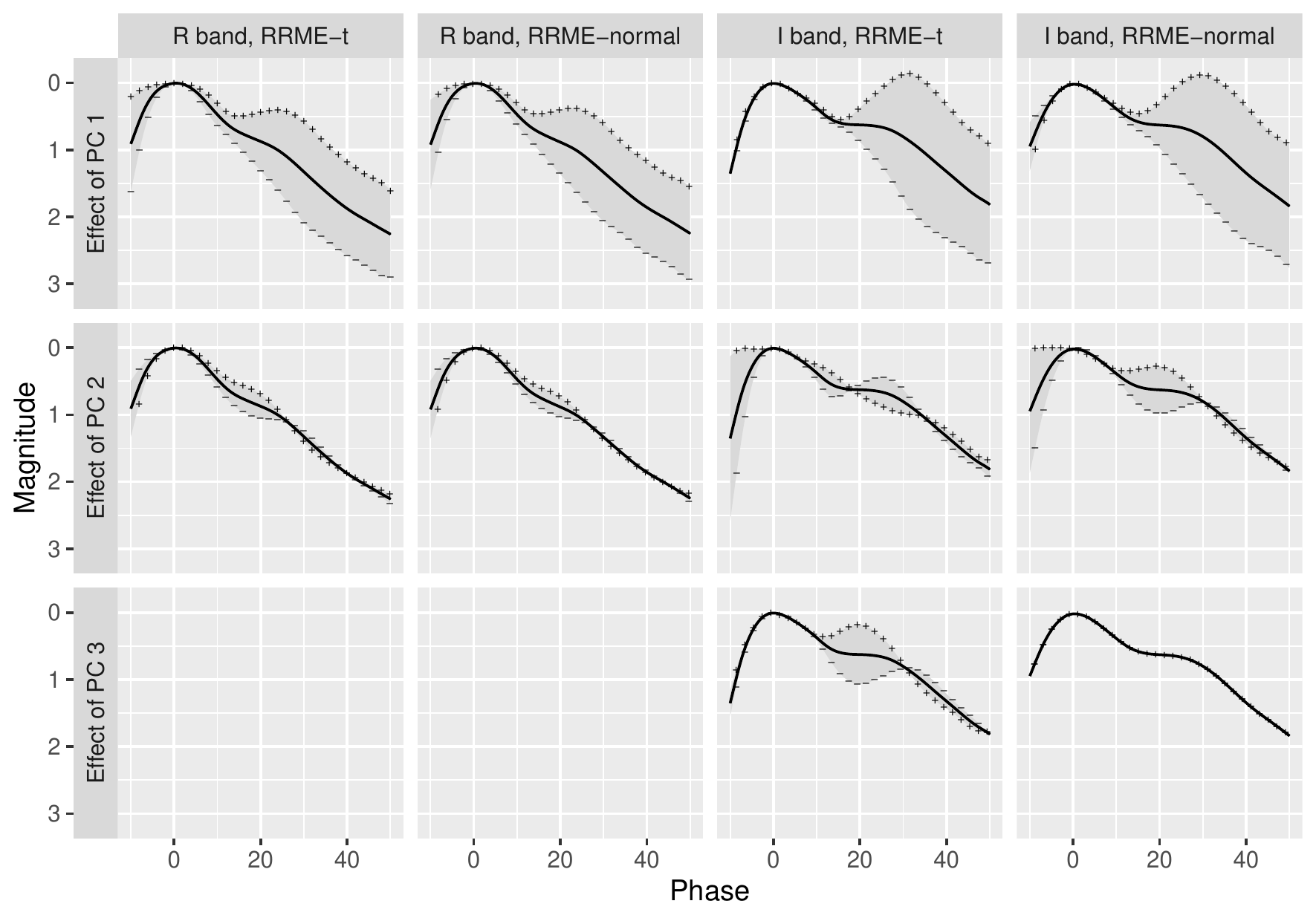}
\caption[.]{
The estimated mean functions and effects of functional principal
components for the type Ia supernova light curve example.
Each panel shows the effect of one PC by plotting the estimated mean function (solid lines) and $\pm$ 2 $\times$ standard deviation of the
PC score $\times$ PC function ("+" or "-" signs).}
\label{fig:mean+-pc} 
\end{figure}

Figure~\ref{fig:mean+-pc} shows the estimated mean functions and the
effects of the estimated PCs using the RRME-t model and the RRME-normal model. For the R band data, we can see that both models provide similar estimates of the mean functions and the PCs.
For the I band data, both models give similar estimates of the mean functions, the first PC functions and the variances of the corresponding PC scores. However, the estimates are very different for the second and the third PCs. The second PC from the RRME-t model captures the changes of the declining rates before and after the second peak, describing the intrinsic variability of the second peak location of normal
SNeIa. Note that this feature is completely missed by the RRME-normal model. 
The third PC from the RRME-t model is similar to the second PC from the RRME-normal model, which adjusts the overall magnitude around the secondary peak. The contribution from the third PC from the RRME-normal model is negligible as the variance of the third PC score is near zero.

\section{Discussion}

In this paper, we have developed a robust method for jointly modelling sparsely observed paired functional data.
Our method is a novel combination of the reduced-rank mixed effects (RRME) model and the scale mixture of normal (SMN) distributions. The EM algorithm provides an efficient computational approach for fitting the model. Our method performs well in simulation studies and in application to a real dataset. We found that our method is not sensitive to two popular choices of the fat-tailed distributions for PC scores and error terms; 
using the generalized Student $t$ or the slash distributions gives similar results.

While our method is presented for paired functional data, it can be extended easily to model multivariate functional data. One simply needs to replace the multivariate normal distribution used in the scale mixture by a higher dimensional one for modelling the PC score vectors from multiple functions. The EM algorithm and its implementation are extendable in a straightforward fashion.

One limitation of our methodology is that the same scale variable $u$ is used for the PC scores and the error terms in the model. It would be more natural to use different scale variable for the PC scores and for the error terms, but the resulting joint Student $t$ distribution of the observed data would no longer be a scale mixture of normal distributions. Our restricted formulation is necessary to ensure that the E step of the EM algorithm has an analytical form. Relaxing this restriction does not change the theoretical framework, but a more complicated and computationally more expensive algorithm, such as the Monte Carlo EM algorithm, is needed for computation.
Implementation of this extension of the methodology is left for future research.

\bibliographystyle{apalike}

\bibliography{reference}

\section*{Appendix A}
\label{sub:em}

\subsection*{A.1. E step}

In the E step, we need to obtain the penalized conditional expected
log-likelihood \eqref{eq:pen-expec-lik}. Based on
\eqref{eq:comp-log-lik}, we need only to evaluate the conditional expectation of $u_i, \log u_i, u_i\balpha_i, u_i\bbeta_i, u_i \balpha_i \balpha_i\trans, u_i\bbeta_i\bbeta_i\trans, u_i\balpha_i\bbeta_i\trans, u_i\bbeta_i\balpha_i\trans$ and $u_i\, (\balpha_i\trans,\bbeta_i\trans) \bSigab^{-1}
  (\balpha_i\trans,\bbeta_i\trans) \trans$ given observations $\bY_i$ and $\bZ_i$, for $i\in\{1,\ldots,n\}$.

We first obtain the conditional expectation of $U_i$ and $\log\, U_i$, denoted as $\hat{u}_i$ and $\widehat{\log\, u_i}$,
by calculating the conditional density of $U_i$ given  $(\bY_i, \bZ_i)$, $\pi(u_i|\bY_i, \bZ_i)$. We have that $\pi(u_i|\bY_i, \bZ_i) \propto \pi(u_i) f(\bY_i,\bZ_i|u_i)$, 
where $\pi(u_i)$ is the marginal density of $U_i$,
and $f(\bY_i,\bZ_i\vert u_i)$ is the conditional density of $(\bY_i,\bZ_i)$ given $U_i = u_i$. Note that $(\bY_i,\bZ_i)$ given $U_i=u_i$ follows a multivariate normal distribution
\begin{eqnarray*}\label{eq:data|u}
\begin{pmatrix}\bY_i\\ \bZ_i\end{pmatrix}\bigg\vert u_i
\sim \mathcal{N} \left(
\begin{pmatrix}\bBy_i \bthmu\\ \bBz_i\bthnu\end{pmatrix},
\frac{1}{u_i} \bSigma_i
\right),
\end{eqnarray*}
where
\begin{align*}\label{eq:var-data|u}
\bSigma_i 
& = \begin{pmatrix}
\bBy_i\bThf \bDa \bThf\trans {\bBy_i}\trans  + \sigma_\epsilon^2
\mbf{I}_{n_i} 
& \bBy_i\bThf \bC \bThg\trans {\bBz_i}\trans\\
\bBz_i\bThg \bC\trans \bThf\trans {\bBy_i}\trans 
& \bBz_i\bThg \bDb \bThg\trans {\bBz_i}\trans  + \sigma_\xi^2
\mbf{I}_{m_i}
\end{pmatrix}.
\end{align*}
In Supplementary Material, Section B, we derive the conditional expectation of $U_i$ and $\log U_i$ given $(\bY_i, \bZ_i)$ in the RRME-t and RRME-slash models.

Next, we evaluate the conditional expectation of $\balpha_i$ and $\bbeta_i$. 
Conditioned on $U_i=u_i$, we have that 
\begin{eqnarray*}\label{eq:cov-re.data|u}
\cov \left(
\begin{pmatrix}\balpha_i\\ \bbeta_i\end{pmatrix},
\begin{pmatrix}\bY_i\\ \bZ_i\end{pmatrix}
\bigg\vert U_i=u_i 
\right)
= \frac{1}{u_i} \bSigabi = \frac{1}{u_i} \begin{pmatrix} \bDa & \bC \\ \bC\trans & \bDb \end{pmatrix}
\begin{pmatrix} \mbf{\Theta}_\mbf{f}\trans {\bBy_i}\trans & \mbf{0} \\ \mbf{0} & \mbf{\Theta}_\mbf{g}\trans {\bBz_i}\trans \end{pmatrix}.
\end{eqnarray*}
Let $\bar\bmu_i = \bSigabi \bSigma_{i}^{-1} (\bY_i - \bBy_i \bthmu)$, $\bar\bnu_i = \bSigabi \bSigma_{i}^{-1} (\bZ_i - \bBz_i \bthnu )$, and $\bar\bSigma_i = \bSigab - \bSigabi \bSigma_{i}^{-1}\bSigabi\trans$. We obtain that
\begin{eqnarray}\label{eq:re|u,data}
(\balpha_i\trans, \bbeta_i\trans)\trans \bigg\vert 
U_i = u_i, \bY_i,\bZ_i
\quad \sim \quad \mathcal{N} \biggl(
(\bar\bmu_i\trans, \bar\bnu_i\trans)\trans,
\frac{1}{u_i} \bar\bSigma_i \biggr).
\end{eqnarray}
Therefore,
$\hat\balpha_i = E(\balpha_i|\bY_i,\bZ_i)  = \bar\bmu_i\quad\text{and}\quad
\hat\bbeta_i = E(\bbeta_i|\bY_i,\bZ_i)  = \bar\bnu_i.$

Note that $\bar{\mu}_i$ and $\bar{\nu}_i$ are independent of $U_i$. Using \eqref{eq:re|u,data} and the law of iterated
expectation,  we obtain that
\begin{align*}
(\widehat{u_i\balpha_i}\trans, \widehat{u_i\bbeta_i}\trans)\trans
&=
E\bigl\{U_i \,(\balpha_i\trans, \bbeta_i\trans)\trans
\big\vert \bY_i, \bZ_i \bigr\}\\
&=
E \bigl\{ U_i E  \bigl[ (\balpha_i\trans, \bbeta_i\trans)\trans
\big\vert U_i=u_i,\bY_i, \bZ_i \bigr] \big\vert \bY_i, \bZ_i\bigr\} \\
&= \hat{u}_i  (\bar\bmu_i\trans, \bar\bnu_i\trans)\trans
\end{align*}
and
\begin{align*}
\begin{pmatrix} \widehat{u_i\balpha_i\balpha_i\trans}
& \widehat{u_i\balpha_i\bbeta_i\trans}\\
\widehat{u_i\bbeta_i\balpha_i\trans}
& \widehat{u_i\bbeta_i\bbeta_i\trans}
\end{pmatrix}
&=
E\biggl\{u_i\, 
 (\balpha_i\trans, \bbeta_i\trans)\trans
 (\balpha_i\trans, \bbeta_i\trans)
\bigg\vert \bY_i, \bZ_i \biggr\}\\
&= \hat{u}_i
(\bar\bmu_i\trans, \bar\bnu_i\trans)\trans
(\bar\bmu_i\trans, \bar\bnu_i\trans)
+ \bar\bSigma_i.
\end{align*}
Therefore,
\begin{eqnarray*}
E\biggl\{u_i\, (\balpha_i\trans,\bbeta_i\trans) \bSigab^{-1}
 (\balpha_i\trans,\bbeta_i\trans)\trans
\bigg\vert \bY_i, \bZ_i \biggr\}
= \trace \biggl[ \bSigab^{-1}\, 
\biggl\{\hat{u}_i (\bar\bmu_i\trans, \bar\bnu_i\trans)\trans
(\bar\bmu_i\trans, \bar\bnu_i\trans)
+ \bar\bSigma_i \biggr\}\biggr].
\end{eqnarray*}

\subsection*{A.2. M step}

In the M step, we update the parameters by minimizing the penalized
expected log-likelihood obtained in the E step.
Since the parameters are well separated in the objective
function, we update the estimates of the parameters 
sequentially in the following order: 
(1) $\sigma_\epsilon^2$ and $\sigma_\xi^2$,
(2) $\bthmu$ and $\bthnu$, 
(3) $\bThf$, $\bThg$, and $\bSigab$, 
(4) $\gamma$. 
While we update one set of parameters, the other parameters are fixed at the values
obtained from the previous iteration.
Details are given below. 

1. Update the estimates of $\sigma_\epsilon^2$ and $\sigma_\xi^2$ as the following,
\begin{align*}
\hat\sigma_\epsilon^2 
&= \frac{1}{\sum_{i=1}^n n_i}\sum_{i=1}^n
\{\hat{u}_i(\bY_i-\bBy_i\bthmu - \bBy_i\bThf\hat\balpha_i)\trans
(\bY_i-\bBy_i\bthmu - \bBy_i\bThf\hat\balpha_i)\\
&\qquad + {\rm tr}(\bBy_i\bThf\bar\bSigma_{i,\balpha\balpha}\mbf{\Theta}_{\mbf{f}}\trans  {\bBy_i}\trans )\}, \\
\hat\sigma_\xi^2 
&= \frac{1}{\sum_{i=1}^n m_i}\sum_{i=1}^n
\{\hat{u}_i(\bZ_i-\bBz_i\bthnu - \bBz_i\bThg\hat\bbeta_i)\trans
(\bZ_i-\bBz_i\bthnu - \bBz_i\bThg\hat\bbeta_i)\\
&\qquad + {\rm tr}(\bBz_i\bThg\bar\bSigma_{i,\bbeta\bbeta}\mbf{\Theta}_{\mbf{g}}\trans  {\bBz_i}\trans )\}.
\end{align*}

2. Update the estimates of $\bthmu$ and $\bthnu$ as the following,
\begin{eqnarray*}
\hat{\btheta}_{\bmu} 
= \biggl(\sum_{i=1}^n \hat{u}_i{\bBy_i}\trans \bBy_i
+ n\sigma_\epsilon^2 \lambda_\mu \bOmega \biggr)^{-1}
\sum_{i=1}^n {\bBy_i}\trans (\hat{u}_i \bY_i - \bBy_i\bThf\widehat{u_i\balpha_i}),
\\
\hat{\btheta}_{\bnu}
 = \biggl(\sum_{i=1}^n \hat{u}_i {\bBz_i}\trans \bBz_i
+ n\sigma_\xi^2 \lambda_\nu \bOmega \biggr)^{-1}
\sum_{i=1}^n {\bBz_i}\trans (\hat{u}_i  \bZ_i - \bBz_i\bThg\widehat{u_i\bbeta_i}).
\end{eqnarray*}

3. To estimate $\bThf,\ \bThg$ and $\bSigab$, 
we first update the estimate of  $\bSigab$ by minimizing the objective function
\begin{eqnarray*}
n \,\log \,|\bSigab| + \trace \left\{\bSigab^{-1} 
\sum_{i=1}^n \begin{pmatrix} \widehat{u_i\balpha_i\balpha_i\trans}
& \widehat{u_i\balpha_i\bbeta_i\trans}\\
\widehat{u_i\bbeta_i\balpha_i\trans}
& \widehat{u_i\bbeta_i\bbeta_i\trans}
\end{pmatrix}\right\}, 
\end{eqnarray*}
and the minimizer is
\begin{eqnarray*}
\hat\bSigma_{\balpha\bbeta}^* = \frac{1}{n} \sum_{i=1}^n 
\begin{pmatrix} \widehat{u_i\balpha_i\balpha_i\trans}
& \widehat{u_i\balpha_i\bbeta_i\trans}\\
\widehat{u_i\bbeta_i\balpha_i\trans}
& \widehat{u_i\bbeta_i\bbeta_i\trans}
\end{pmatrix}
=
\begin{pmatrix}
\widehat{\bV}_{\alpha\alpha} & \widehat{\bV}_{\alpha\beta}\\
\widehat{\bV}_{\beta\alpha} & \widehat{\bV}_{\beta\beta}\\
\end{pmatrix}.
\end{eqnarray*}

Denote $\bThf^*= (\theta_{f 1}^*, \theta_{f 2}^*,
\dots, \theta_{f k_\alpha}^*)$ and $\bThg^* = (\theta_{g 1}^*, \dots, \theta_{g k_\beta}^*)$ as the minimizers of the penalized expected log-likelihood, we update their columns sequentially. 
The update formulas are, for $1\leq j\leq k_\alpha$,
\begin{align*}
\hat\theta_{f j}^*
&= \biggl(\sum_{i=1}^n \widehat{u_i\alpha_{ij}^2} {\bBy_i}\trans \bBy_i
+n\sigma^2_\epsilon \var(\alpha_j) \lambda_f \bOmega \biggr)^{-1} \\
&\qquad \times \sum_{i=1}^n {\bBy_i}\trans \biggl\{
(\bY_i-\bBy_i\bthmu)\widehat{u_i\alpha_{ij}}
- \sum_{l \not = j} \bBy_i\theta_{f l}
\,\widehat{u_i \alpha_{il}\alpha_{ij}}\biggr\},
\end{align*}
and for $1\leq j \leq k_\beta$,
\begin{align*}
\hat\theta_{g j}^*
&= \biggl(\sum_{i=1}^n \widehat{u_i\beta_{ij}^2} {\bBz_i}\trans \bBz_i
+n\sigma^2_\xi \var(\beta_j)\lambda_g \bOmega \biggr)^{-1}\\
&\qquad \times
\sum_{i=1}^n {\bBz_i}\trans \biggl\{
(\bZ_i-\bBz_i\bthnu)\widehat{u_i\beta_{ij}}
- \sum_{l \not =  j} \bBz_i\theta_{g l}
\,\widehat{u_i\beta_{il}\beta_{ij}}\biggr\}.
\end{align*}

Note that $\bThf^*$ and $\bThg^*$ may not be orthonormal and $\widehat{\bV}_{\alpha\alpha}, \widehat{\bV}_{\beta\beta}$ may not be diagonal with decreasing values. For identifiability, we perform the
following singular value decompositions to get the estimates $\widehat\bThf$, $\widehat\bThg$ and $\widehat\bSigab$ such that
$
\widehat\bThf^*\widehat{\bV}_{\balpha\balpha}\widehat\bThf^{*\rm T}
= \widehat\bThf \widehat{\mbf{D}}_{\balpha} \widehat\bThf\trans $ and $
\widehat\bThg^*\widehat{\bV}_{\bbeta\bbeta}\widehat\bThg^{*\rm T}
= \widehat\bThg \widehat{\mbf{D}}_{\bbeta} \widehat\bThg,
$
where $\widehat\bThf$ and $\widehat\bThg$ are orthogonal matrices, 
$\widehat{\mbf{D}}_{\balpha}$ and $\widehat{\mbf{D}}_{\bbeta}$ are diagonal
matrices with decreasing diagonal elements. 
Moreover, 
\begin{eqnarray*}
&\begin{pmatrix}
\widehat\bThf^* & \bf{0} \\
\bf{0} & \widehat\bThg^* 
\end{pmatrix}
\widehat\bSigab^*
\begin{pmatrix}
(\widehat\bThf^*)\trans & \bf{0} \\
\bf{0} & (\widehat\bThg^*)\trans
\end{pmatrix}
= 
\begin{pmatrix} \widehat\bThf & \bf{0} \\ \bf{0} & \widehat\bThg\\ \end{pmatrix}
\widehat\bSigab
\begin{pmatrix} \widehat\bThf\trans & \bf{0} \\ \bf{0} & \widehat\bThg\trans\\ \end{pmatrix},\\
&\quad \text{ where }\quad \widehat\bSigab=\begin{pmatrix}
\widehat{\mbf{D}}_{\balpha}&\widehat{\bC}\\ \widehat{\bC}\trans&\widehat{\mbf{D}}_{\bbeta}
\end{pmatrix}.
\end{eqnarray*}
Therefore, 
$\widehat{\bC} = \widehat\bThf \trans\widehat\bThf^* \widehat\bSigab^*
\widehat{\mbf{\Theta}}_\mbf{g}^{*\rm T} \widehat\bThg$.

4. We  minimize $-2\sum_{i=1}^n E_{U_i}[\log\ p_u(u_i)|\bY_i,\bZ_i;\Xi^{(l)}]$ to update  the estimate of $\gamma$, where $\Xi^{(l)}$ is the current estimates of the parameters.

\end{document}